\documentclass[pra,twocolumn,amsmath,amssymb,aps,superscriptaddress]{revtex4}
\newtheorem{thm}    {Theorem}

\def\Tr{\mathop{\rm Tr}\nolimits}

\def\rank{\mathop{\rm rank}\nolimits}

\def\real{\mathbb{R}}

\def\complex{\mathbb{C}}

\def\SU{\mathop{\rm SU}\nolimits}

\def\sgn{\mathop{\rm sgn}\nolimits}

\def\Label#1{\label{#1}\ [\ #1\ ]\ }
\def\Label{\label}

\begin{document}
\title{Universal approximation of multi-copy state\\ and universal quantum lossless data compression}
\author{Masahito Hayashi}
\email{hayashi@math.is.tohoku.ac.jp}
\address{Graduate School of Information Sciences, Tohoku University, Aoba-ku, Sendai, 980-8579, Japan}

\begin{abstract}
We have proven that there exists a quantum state approximating any multi-copy state universally when we measure the error by the normalized relative entropy.
While the qubit case was proven by Krattenthaler and Slater 
(IEEE Trans. IT {\bf 46}, 801-819 (2000)), 
the general case has been open for more than ten years.
For a deeper analysis, we have solved the mini-max problem concerning `this approximation error' up to the second order.
Further, we have applied this result to quantum lossless date compression,
and have constructed a universal quantum lossless data compression.
\end{abstract}

\maketitle
\section{Introduction}\Label{s1}
Does there exist a quantum state approximating the $n$-copy state universally?
Such a state does not seem to exist, however, 
if we measure the difference between two states by the normalized relative entropy, such a state exists.
In the classical case, i.e., the non-commutative case,
Clarke and Barron \cite{CB} shown that the mixture state of $n$-fold tensor product states 
approximates all of $n$-fold tensor product states.
More precisely, they proven that the following.
Consider the set of commutative desities 
$\{\rho_\theta |\theta\in \Theta \subset \real^m \}_{\theta}$ on ${\cal H}:=\complex^d$,
and a prior distribution $\mu$ on $\Theta$. 
They shown that
the mixture state $\rho_{w,n}:=\int_\Theta \rho_\theta 
w(\theta) d \theta$
satisfies the relation
\begin{align}
D(\rho^{\otimes n}_\theta \| \sigma_{\mu,n} )
\cong \frac{m}{2}\log \frac{n}{2\pi e} +
\frac{1}{2}\log \det \sqrt{I(\theta)}-\log w (\theta) ,
 \Label{P1}
\end{align}
where
$I(\theta)$ is the Fisher information matrix (See \cite{AM}).
In this paper, we choose the logarithm base $2$.
In particular, Clarke and Barron \cite{CB2} proven that 
the mini-max of the constant term is given when
the prior $w$ is chosen as the Jeffreys' prior 
$\frac{\sqrt{\det I(\theta)}}{\int \sqrt{\det I(\theta')}d \theta'}$.
If we replace the relative entropy by another measure satisfying the axiom of the distance, 
no state satisfies the relation (\ref{P1}).
That is, since the relative entropy does not satisfy the axiom of the distance, 
such a state can be exist.
Concerning the quantum case,
in 1996, Krattenthaler and Slater\cite{KS} shown its quantum extension in the qubit case.
However, the general case has not been shown for more than ten years.
Their paper has not proven the mini-max problem for the constant term completely.

In this paper, we prove the existence of the states $\sigma_n$ on ${\cal H}^{\otimes n}$
satisfying that
\begin{align}
D(\rho^{\otimes n} \| \sigma_{n} )
\cong \frac{d^2-1}{2}\log n + O(1)\Label{P2}
\end{align}
for all of faithful states $\rho$ on ${\cal H}$, i.e., the state $\rho$ belongs to the set ${\cal S}:=\{\rho|\rank \rho=d\}$.
Since the dimension of the state family ${\cal S}$ is $d^2-1$,
the relation (\ref{P2}) can be regarded as a natural quantum extension of (\ref{P1}).
More precisely, we calculate the following mini-max value
\begin{align}
\min_{\{\sigma_n \}}
\sup_{\rho \in {\cal S}}
\lim_{n \to \infty}
\left(D(\rho^{\otimes n} \| \sigma_{n} )
-\frac{d^2-1}{2}\log n\right),\Label{P3}
\end{align}
which is the main result in this paper.
Krattenthaler and Slater\cite{KS} treated the same problem among a restricted class of states $\{\sigma_n\}$ in the qubit case.

The term (\ref{P3}) has the meaning of the approximation in the sense of information processing.
Recently, Hayashi \cite{U-channel} provided universal coding for classical-quantum channel.
In his derivation, the approximating state $\sigma_n$ is used instead of the true state $\rho^{\otimes n}$.
The reason why this replacement well-works seems that this type information-sense approximation holds.

In the classical case, this relation is considered as the asymptotic redundancy of the prefixed variable-length lossless data compression\cite{CB,CB2}.
Any distribution corresponds to the prefixed variable-length data compression through Kraft inequality,
and the minimum average length is the Shannon entropy\cite{HK}.
When the applied code corresponds to the distribution $\sigma_{\mu,n}$,
the relation (\ref{P1}) expresses the asymptotic redundancy when the true state is $\rho^{\otimes n}$.
This fact means that good information-sense approximation implies a small redundancy in the prefixed variable-length lossless data compression.
That is, the relation (\ref{P1}) guarantees the existence of 
universal variable-length lossless data compression in the classical case\cite{CB,CB2,Ly,Da}.

In the quantum case, it is known that there exists 
universal fixed-length data compression\cite{JH,Expo-s}.
However, the quantum analogue of variable-length lossless data compression is not simple\cite{BF}
In order to determine the size of the memory storing the quantum information,
we have to measure the length.
This measurement demolishes the quantum state while the degree of the state demolition can be reduced 
to infinitesimal\cite{HayaMa}.
Hence, it is impossible to interpret the length of the quantum case in the same way.
Therefore, we have to prepare the memory whose size is the maximum length of the given code
in the framework of variable-length lossless data compression\cite{KI}.
In order to avoid this problem, we regard the length of code as the energy
and treat the average energy because the energy is not necessarily to be measured.
In this paper, we formulate quantum variable-length lossless data compression in this way,
and explain that the relative entropy $D(\rho^{\otimes n}\|\sigma_n)$ expresses the asymptotic redundancy.

This paper is organized as follows.
In section \ref{s2}, we prepare a basic knowledge of representation theory,
and show (\ref{P2}). In section \ref{s3}, using the relations given in section \ref{s2}, we calculate (\ref{P3}). In section \ref{s4}, we numerically calculate (\ref{P3}) in the qubit case,
and compare the result by Krattenthaler and Slater \cite{KS}.
In section \ref{s5}, we treat quantum variable-length lossless data compression, and apply our result to this topic.

\section{Representation theory and approximation of multi-copy state}\Label{s2}
In order to treat information-sense approximation,
we focus on the dual representation on the $n$-fold tensor product space by the 
the special unitary group $\SU(d)$ and the $n$-th symmetric group $S_n$.
(Christandl\cite{Christandl} contains
a good survey of representation theory for quantum information.)
For this purpose, we focus on Young diagram and type.
When the vector of integers $\vec{n}=(n_1,n_2, \ldots, n_d)$ satisfies the condition $n_1 \ge n_2 \ge \ldots \ge n_d\ge 0$ and $\sum_{i=1}^d n_i=n$,
the vector $\vec{n}$ is called the Young diagram (frame) with the size $n$ and the depth $d$, 
their set is denoted by $Y_n^d$.
The number of these sets is evaluated by
\begin{align}
|Y_n^d| \le (n+1)^{d-1}\Label{9}.
\end{align}
Since 
the sets $\{(n_{s(i)})|\vec{n} \in T_n^d\}$ 
and 
$\{(n_{s'(i)})|\vec{n} \in T_n^d\}$ are distinct for any $s \neq s'\in S_d$,
the relation $|\{\vec{n}|\sum_i n_i=n\}|\cong \frac{n^{d-1}}{(d-1)!}$
implies the asymptotic behavior of the cardinality $|Y_n^d|$:
\begin{align}
|Y_n^d|
\cong 
\frac{n^{d-1}}{d! (d-1)!}
\Label{9a}.
\end{align}
Using the Young diagram, we can characterize the irreducible decomposition of the dual 
representation of $\SU(d)$ and $S_n$:
\begin{align*}
{\cal H}^{\otimes n}=\bigoplus_{\vec{n}\in Y_n^d} {\cal U}_{\vec{n}} \otimes {\cal V}_{\vec{n}},
\end{align*}
where ${\cal U}_{\vec{n}}$ is the irreducible representation space of $\SU(d)$ characterized by $\vec{n}$, and
${\cal V}_{\vec{n}}$ is the irreducible representation space of $n$-th symmetric group $S_n$ characterized by $\vec{n}$.
Here, we denote the representation of $n$-th symmetric group $S_n$ by $V:  S_n \ni s \mapsto V_s$.
According to Weyl's dimension formula,
the dimension ${\cal U}_{\vec{n}}$ has the expression
\begin{align}
\dim {\cal U}_{\vec{n}} =
\prod_{i<j}\frac{n_i-n_j+j-i}{j-i}< n^{\frac{d(d-1)}{2}}
\Label{11}.
\end{align}
for any $\vec{n}\in Y_n^d$, 
That is,
for a given probability distribution 
$\vec{p}=(p_1, \ldots, p_d)$ on $\{1, \ldots, d\}$
with the condition $p_1 > p_2 > \cdots > p_d$,
when $(p_1, \ldots, p_d)=(\frac{n_1}{n}, \ldots, \frac{n_d}{n})$,
\begin{align}
\dim {\cal U}_{\vec{n}} 
\cong
\frac{\prod_{i<j}(p_i-p_j)}
{2^{d-1} 3^{d-2} \cdots (d-1)^{1}}n^{\frac{d(d-1)}{2}}
\Label{a11}.
\end{align}
Then, we denote the projection to the subspace ${\cal U}_{\vec{n}} \otimes {\cal V}_{\vec{n}}$ by $I_{\vec{n}}$, and define the following.
\begin{align}
\rho_{\vec{n}}&:= \frac{1}{\dim {\cal U}_{\vec{n}} \otimes {\cal V}_{\vec{n}}} I_{\vec{n}}
\nonumber \\
\sigma_{U,n}&:= \sum_{\vec{n}\in Y_n^d} \frac{1}{|Y_n^d|} \rho_{\vec{n}}.\Label{15}
\end{align}
Any state $\rho$ and any Young diagram $\vec{n}\in Y_n^d$ satisfy that
\begin{align*}
\dim {\cal U}_{\vec{n}} \rho_{\vec{n}} 
\ge I_{\vec{n}} \rho^{\otimes n} I_{\vec{n}}.
\end{align*}
Thus, (\ref{9}), (\ref{11}), and (\ref{15}) yield 
the inequality
\begin{align}
(n+1)^{\frac{(d+2)(d-1)}{2}}
 \sigma_{U,n} \ge \rho^{\otimes n}.\Label{ineq-2}
\end{align}
Since $ \sigma_{U,n}$ is commutative with $\rho^{\otimes n}$,
we have
\begin{align*}
\log (n+1)^{\frac{(d+2)(d-1)}{2}}+
\log  \sigma_{U,n} \ge \log \rho^{\otimes n}.
\end{align*}
Thus, we obtain
\begin{align*}
D(\rho^{\otimes n}\| \sigma_{U,n})
=& \Tr \rho^{\otimes n}(
\log \rho^{\otimes n}
-\log  \sigma_{U,n} ) \\
\le &\frac{(d+2)(d-1)}{2} \log (n+1).
\end{align*}
Therefore,
the state $\sigma_{U,n}$ universally approximates the state $\rho^{\otimes n}$
in the sense of the normalized quantum relative entropy:
\begin{align*}
\frac{1}{n}D(\rho^{\otimes n}\| \sigma_{U,n})
\to 0.
\end{align*}

Now, we calculate the value $D(\rho^{\otimes n}\| \sigma_{U,n})$ more precisely.
In the following calculation, we assume that
\begin{align*}
\rho(\vec{p})=\sum_{i=1}^d p_i |i \rangle \langle i |,
\end{align*}
where $\vec{p}=(p_1, \ldots, p_d)$ is a probability distribution on $\{1, \ldots, d\}$, and
$\{|i\rangle\}_{i=1}^d$ is the orthonormal basis of ${\cal H}$.
In this calculation, it is essential to calculate 
the average of the random variable $\log |Y^d_n|
+\log \dim {\cal U}_{\vec{n}} 
+\log \dim {\cal V}_{\vec{n}} 
$ under the distribution $Q_{\vec{p}}(\vec{n}):=
\Tr \rho(\vec{p})^{\otimes n} I_{\vec{n}}$.
In order to treat $\sum_{\vec{n}\in Y_n^d} Q_{\vec{p}}(\vec{n}) 
\log \dim {\cal V}_{\vec{n}}$ asymptotically, 
Matsumoto and Hayashi \cite{MaHa07} introduced the quantity $\frac{n!}{\vec{n}!} :=\frac{n!}{n_1! n_2! \cdots n_d!}$.
In their Appendix D, they shown that
\begin{align}
&\sum_{\vec{n}\in Y_n^d} Q_{\vec{p}}(\vec{n}) 
(\log \dim {\cal V}_{\vec{n}} - \log \frac{n!}{\vec{n}!}) \nonumber \\
\cong &
\sum_{s \in S_d} 
\frac{\sgn (s) \prod_i  p_i^{\delta_{s(i)}}  }
{\prod_{i<j}(p_i-p_j)}
\log 
\frac{\prod_{i<j}(p_i-p_j)}
{ \prod_i  p_i^{\delta_{s(i)}}  } \nonumber\\
=&
\log \prod_{i<j}(p_i-p_j)
-\sum_{s \in S_d} 
\frac{\sgn (s) \prod_i  p_i^{\delta_{s(i)}}  }
{\prod_{i<j}(p_i-p_j)}
\log 
\prod_i  p_i^{\delta_{s(i)}}  
\Label{6-2-1} ,
\end{align}
where
$\delta_i :=d-i$ and we applied the formula
$\sum_{s \in S_d} \sgn (s) \prod_i  p_i^{\delta_{s(i)}} =\prod_{i<j}(p_i-p_j)$.
In their Appendix C, they calculated 
$\sum_{\vec{n}\in Y_n^d} Q_{\vec{p}}(\vec{n})$ as
\begin{align}
&\sum_{\vec{n}\in Y_n^d} Q_{\vec{p}}(\vec{n}) 
\log \frac{n!}{\vec{n}!}  \nonumber \\
\cong & H(\vec{p})n - \frac{d-1}{2}\log n -\frac{d-1}{2}\log 2 \pi e
- \frac{1}{2}\sum_{i}\log p_i.\Label{6-2-2}
\end{align}
The combination of (\ref{9a}), (\ref{a11}), (\ref{6-2-1}), and (\ref{6-2-2})
yields that
\begin{align*}
&\sum_{\vec{n}\in Y_n^d} {\rm P}(\vec{n}) 
(\log |Y^d_n|+\log \dim {\cal U}_{\vec{n}} +\log \dim {\cal V}_{\vec{n}} )\\
\cong &
H(\vec{p})n +
\log \prod_{i<j}(p_i-p_j)
-\sum_{s \in S_d} 
\frac{\sgn (s) \prod_i  p_i^{\delta_{s(i)}}  }
{\prod_{i<j}(p_i-p_j)}
\log 
\prod_i  p_i^{\delta_{s(i)}}  \\
&- \frac{d-1}{2}\log n -\frac{d-1}{2}\log 2 \pi e 
- \frac{1}{2}\sum_{i}\log p_i \\
& +\log \frac{n^{d-1}}{d! (d-1)!}
+\log \frac{n^{d(d-1)}\prod_{i<j}(p_i-p_j)}
{2^{d-1} 3^{d-2} \cdots (d-1)^{1}}\\
\cong &
H(\vec{p})n +
\frac{d^2-1}{2}\log n
-\frac{d-1}{2}\log 2 \pi e \\
&-\log 2^{d-1} 3^{d-2} \cdots (d-1)^{1} \\
&-\log d! (d-1)! 
+2 \log \prod_{i<j}(p_i-p_j) \\
&-\sum_{s \in S_d} 
\frac{\sgn (s) \prod_i  p_i^{\delta_{s(i)}}  }
{\prod_{i<j}(p_i-p_j)}
\log \prod_i  p_i^{\delta_{s(i)}}  
- \frac{1}{2}\sum_{i}\log p_i\\
\cong &
H(\vec{p})n +
\frac{d^2-1}{2}\log n
+C_d-\log d! (d-1)! +C(\vec{p}),
\end{align*}
where 
\begin{align*}
C_d :=&-\frac{d-1}{2}\log 2 \pi e
-\log 2^{d-1} 3^{d-2} \cdots (d-1)^{1} \\
C(\vec{p}) :=& 
-\sum_{s \in S_d} 
\frac{\sgn (s) \prod_i  p_i^{\delta_{s(i)}}  }
{\prod_{i<j}(p_i-p_j)}
\log \prod_i  p_i^{\delta_{s(i)}}  \\
&+2 \log \prod_{i<j}(p_i-p_j)
- \frac{1}{2}\sum_{i}\log p_i.
\end{align*}
Hence, 
\begin{align*}
&D(\rho^{\otimes n}\| \sigma_{U,n})
= -\Tr 
\rho^{\otimes n}\log  \sigma_{U,n}-nH(\rho)\\
\cong &
\frac{d^2-1}{2}\log n
+C_d-\log d! (d-1)!+C(\vec{p}).
\end{align*}

\section{Mini-max problem}\Label{s3}
From the discussion of the above section,
it is possible to reduce the asymptotic approximation error $D(\rho^{\otimes n}\| \sigma_{U,n})$ 
to $\frac{d^2-1}{2}\log n$ universally.
In this section, we treat the mini-max problem concerning the constant term, which is the second order term.
The following is the main theorem.
\begin{thm}
We obtain the following mini-max value:
\begin{align}
&\min_{\{\sigma_n\}} \sup_{\rho\in {\cal S}} \lim_{n \to \infty}
\left(D(\rho^{\otimes n}\| \sigma_{n})- \frac{d^2-1}{2}\log n\right)\nonumber\\
=&
C_d 
+\log \int_{Y^d} e^{C(\vec{p})} d \vec p,
\Label{6-2-3}
\end{align}
where $Y^d:=\{\vec{p}=(p_1,p_2,\ldots,p_{d-1}, 1-p_1-\ldots-p_{d-1})|
p_1>p_2>\ldots> p_{d-1}>1-p_1-\ldots-p_{d-1}>0\}$, and 
$d \vec p:=d p_1d p_2 \ldots p_{d-1}$.
The above mini-max value
is realized when we choose the mixture state
$\sigma_{J,n}:=
\sum_{\vec{n}\in Y^d_n}
J_n(\vec{n})\rho_{\vec{n}}$
with the distribution $J_n(\vec{n}):=\frac{e^{C(\frac{\vec{n}}{n})}}{
\sum_{\vec{n}'\in Y^d_n}e^{C(\frac{\vec{n}'}{n})}}$.
This mini-max value is also attained by 
the mixture state
$
\tilde{\sigma}_{J,n}:=
\int_{Y^d} 
\overline{\rho(\vec{p})^{\otimes n}}
J(\vec{p}) d \vec{p}$, where
$\overline{\rho(\vec{p})^{\otimes n}}$
is the mixture concerning the invariant measure $\mu$ on $\SU(d)$, i.e., 
$\int_{\SU(d)} (U \rho(\vec{p}) U^{\dagger})^{\otimes n} \mu(d U)$.
\end{thm}
Since the Jeffreys' prior gives the mini-max solution in the classical case,
the distribution $J(\vec{p})$ on $Y^d$ can be regarded as a quantum extension of 
Jeffreys' prior.

Since the state $\rho^{\otimes n}$ is invariant for permutation,
$D(\rho^{\otimes n}\| \sigma_{n})=
-\Tr [\rho^{\otimes n} \log \sigma_{n}]- n H(\rho)$ is equal to 
$-\Tr [\rho^{\otimes n} \log V_s \sigma_{n} V_s^{\dagger}]- n H(\rho)$ for any $s \in S_n$.
Thus, the operator convexity of the function $x \mapsto -\log x$ implies
\begin{align*}
& -\Tr [\rho^{\otimes n} \log \sigma_{n}]- n H(\rho) \\
=& 
\sum_{s \in S_n}\frac{1}{|S_n|}
\Tr [\rho^{\otimes n} (-\log V_s \sigma_{n} V_s^{\dagger})]- n H(\rho)\\
\ge &
-\Tr [\rho^{\otimes n} \log \sum_{s \in S_n}\frac{1}{|S_n|} V_s \sigma_{n} V_s^{\dagger}]- n H(\rho).
\end{align*}
Since the state $\sum_{s \in S_n}\frac{1}{|S_n|} V_s \sigma_{n} V_s^{\dagger}$ is 
invariant for the action of $S_n$,
we can restrict our states $\sigma_n$ in (\ref{6-2-3}) to permutation invariant states.
Due to the unitary covariance and
the operator convexity of the function $x \mapsto - \log x$,
the invariant measure $\mu$ on $\SU(d)$ satisfies that
\begin{align*}
& \sup_{\rho} \lim_{n \to \infty}
\left(D(\rho^{\otimes n}\| \sigma_{n})- \frac{d^2-1}{2}\log n\right)\\
= &
\sup_{\vec{p}} \lim_{n \to \infty}
\sup_{U\in SU(d)}
\left(D(\rho(\vec{p})^{\otimes n}\| U^{\otimes n} \sigma_{n} (U^{\otimes n})^{\dagger}
)- \frac{d^2-1}{2}\log n\right)\\
= &
\sup_{\vec{p}} 
\sup_{U\in \SU(d)}
\lim_{n \to \infty}
-\Tr 
\rho(\vec{p})^{\otimes n}
\log  U^{\otimes n} \sigma_{n} (U^{\otimes n})^{\dagger} \\
&\qquad - n H(\vec{p})
- \frac{d^2-1}{2}\log n \\
\ge &
\sup_{\vec{p}} 
\lim_{n \to \infty}
- \int_{\SU(d)}
\Tr 
\rho(\vec{p})^{\otimes n}
\log  U^{\otimes n} \sigma_{n} (U^{\otimes n})^{\dagger}
\mu(d U) \\
&\qquad - n H(\vec{p})
- \frac{d^2-1}{2}\log n \\
\ge &
\sup_{\vec{p}} 
\lim_{n \to \infty}
-\Tr 
\rho(\vec{p})^{\otimes n}
\log  
\left( \int_{\SU(d)}
U^{\otimes n} \sigma_{n} (U^{\otimes n})^{\dagger}\mu(d U)\right)\\
&\qquad  - n H(\vec{p})
- \frac{d^2-1}{2}\log n .
\end{align*}
Since the state $\left( \int_{\SU(d)}
U^{\otimes n} \sigma_{n} (U^{\otimes n})^{\dagger}\mu(d U)\right)$
is invariant for the action of $\SU(d)$,
we can restrict our states $\sigma_n$ in (\ref{6-2-3}) to states invariant for $\SU(d)$.
That is, we obtain 
\begin{align*}
&\inf_{\{\sigma_n\}} \sup_{\rho} \lim_{n \to \infty}
\left(D(\rho^{\otimes n}\| \sigma_{n})- \frac{d^2-1}{2}\log n\right)\\
=&
\inf_{\{P_n\}} \sup_{\rho} \lim_{n \to \infty}
\left(D(\rho^{\otimes n}\| \sigma_{P_n,n})- \frac{d^2-1}{2}\log n\right),
\end{align*}
where $P_n$ is the probability measure on $Y^d_n$ and 
$\sigma_{P_n,n}:=\sum_{\vec{n} \in Y^d_n}
P_n(\vec{n})\rho_{\vec{n}}$.
Since
\begin{align*}
\frac{\sum_{\vec{n}'\in Y^d_n}e^{C(\frac{\vec{n}'}{n})}}{n^{d-1}}
\to
\int_{Y^d} e^{C(\vec{p})} d \vec p,
\end{align*}
we have
\begin{align}
& D(\rho(\vec{p})^{\otimes n}\| \sigma_{P_n,n}) \nonumber\\
\cong & \frac{d^2-1}{2}\log n
+C_d -\log d!(d-1)!
+C(\vec{p})
-\log \frac{P_n(\vec{n})}{\frac{1}{|Y^d_n|}}
\nonumber \\
\cong & \frac{d^2-1}{2}\log n
+C_d -\log d!(d-1)!
+\log (\int_{Y^d} e^{C(\vec{p}')} d \vec{p}')\nonumber\\
&
+\log J(\vec{p}) 
-\sum_{\vec{n}}Q_{\vec{p}}(\vec{n})
\log P_n(\vec{n})
\frac{n^{d-1}}{d!(d-1)!}
\nonumber\\
\cong &
\frac{d^2-1}{2}\log n
+C_d 
+\log \int_{Y^d} e^{C(\vec{p}')} d \vec{p}' \nonumber\\
&+
\sum_{\vec{n}}Q_{\vec{p}}(\vec{n})
(\log J(\vec{p})-\log P_n(\vec{n})-(d-1)\log n).\Label{6-3-1}
\end{align}

Consider the joint distribution 
$Q_{J,n}(\vec{p},\vec{n}):=J(\vec{p}) Q_{\vec{p}}(\vec{n})$.
The marginal distribution 
$Q_{J,n}(\vec{n})
(=\int_{Y^d} Q_{J,n}(\vec{p},\vec{n})d \vec{p})$ approaches $J_n(\vec{n})$
because the variable $\frac{\vec{n}}{n}$ approaches $\vec{p}$ in probability $Q_{\vec{p}}$\cite{KeylW}
and $J(\vec{p})$ is continuous.
The variable
$\vec{p}$ approaches $\frac{\vec{n}}{n}$ in probability under the conditional distribution $Q_{J,n}(\vec{p}|\vec{n})$.
Then,
\begin{align*}
& \int_{Y^d} 
\log J(\vec{p})
\frac{Q_{\vec{p}}(\vec{n}) J_n(\vec{p})}
{\int_{Y^d}J_n(\vec{p}) {\rm P}_{\vec{p}}(\vec{n}) d \vec{p}} d \vec p \\
\cong &
\log J(\frac{\vec{n}}{n})
=
\log J_n(\vec{n})
+
\log \frac{\sum_{\vec{n}'\in Y^d_n}e^{C(\frac{\vec{n}'}{n})}}{\int_{Y^d} e^{C(\vec{p})} d \vec p}\\
\cong &
\log J_n(\vec{n})
+
(d-1)\log n.
\end{align*}
Only the second group depends on $\vec{p}$ in (\ref{6-3-1}).
The supremum of the second group of (\ref{6-3-1}) concerning $\vec{p}$ 
is evaluated as follows.
\begin{align*}
& \sup_{\vec{p}} 
\sum_{\vec{n}}Q_{\vec{p}}(\vec{n})
(\log J (\vec{p})-\log P_n(\vec{n})- (d-1)\log n) \\
\ge &
\int_{Y^d} 
\sum_{\vec{n}}Q_{\vec{p}}(\vec{n})
(\log J (\vec{p})-\log P_n(\vec{n})-(d-1)\log n) 
J( \vec{p}) d \vec p\\
\cong &
\sum_{\vec{n}}
J_n(\vec{n})
(\log J_n(\vec{n})-\log P_n(\vec{n})) \\
=& D(J_n\| P_n) \ge 0. 
\end{align*}
Hence,
\begin{align*}
& \lim_{n \to \infty} 
\sum_{\vec{n}}
P_n(\vec{n})
(D(\rho(\frac{\vec{n}}{n})^{\otimes n}\| \sigma_{P_n,n})- \frac{d^2-1}{2}\log n) \\
\ge &
C_d 
+\log \int_{Y^d} e^{C(\vec{p})} d \vec p.
\end{align*}
The equality holds when $P_n(\vec{n})=J_n(\vec{n})$.
Therefore,
\begin{align*}
&\min_{\{\sigma_n\}} \sup_{\rho} \lim_{n \to \infty}
\left(D(\rho^{\otimes n}\| \sigma_{n})- \frac{d^2-1}{2}\log n\right) \\
=&
C_d 
+\log \int_{Y^d} e^{C(\vec{p})} d \vec p.
\end{align*}
Note that the existence of the minimum value is proven here.
Since the state $\overline{\rho(\vec{p})^{\otimes n}}$ has the form
$\sum_{\vec{n}\in Y^d_n}
Q_{\vec{p}}(\vec{n})\rho_{\vec{n}}$,
the state $\tilde{\sigma}_{J,n}$ has the form
$\sum_{\vec{n}\in Y^d_n}Q_{J,n}(\vec{n})\rho_{\vec{n}}$.
Because $Q_{J,n}(\vec{n})$ approaches $J_n(\vec{n})$, the state $\tilde{\sigma}_{J,n}$ also attains the mini-max in (\ref{6-2-3}).

\section{Qubit case}\Label{s4}
In the qubit case, $C_2$ and $C(\vec{p})$ are calculated as follows.
\begin{align*}
C_2=&-\frac{1}{2}\log 2\pi e - \log 2\\
C(\vec{p})=& 2\log (p_1-p_2)
-\frac{p_1}{p_1-p_2} \log p_1 \\
&+\frac{p_2}{p_1-p_2} \log p_2
-\frac{1}{2}(\log p_1+\log p_2).
\end{align*}
Then, the term
$\log \int_{Y^2} e^{C(\vec{p})} d \vec p$
is numerically calculated to $-0.50737$.
We obtain
\begin{align*}
\min_{\{\sigma_n\}}
\sup_{\rho \in {\cal S}}\lim_{n \to \infty}(D(\rho^{\otimes n}\|\sigma_n) -\frac{3}{2}\log n )
=-3.5545.
\end{align*}

On the other hand,
Krattenthaler and Slater \cite{KS}
focus on one parameter family $\{\zeta_n(u)|-\infty<u<1\}$,
whose elements are given as the mixtures of $n$-copy state 
for invariant measures $q_u$, which given in (1.7) of \cite{KS}.
They numerically calculated that
\begin{align*}
\min_u \sup_{\rho \in {\cal S}}\lim_{n \to \infty}(D(\rho^{\otimes n}\|\zeta_n(u)) -\frac{3}{2}\log n )
=-2.3956.
\end{align*}
Note that they derived the above value $-1.66050$ with the natural logarithmic base.

This result does not contradict our result because we take the minimum concerning 
all sequence of states $\{\sigma_n\}$ while they take it concerning specific sequences of states $\{\zeta_n(u)\}$.
The difference $- 2.3956-(-3.5545) =1.1589$ expresses the improvement for choosing the optimal approximating state.

\section{Minimization of the average energy}\Label{s5}
In this section, we apply obtained result to quantum variable-length lossless coding.
First, we give a formulation of variable-length lossless coding.
For this purpose, we focus on the Fock space:
\begin{align}
{\cal H}_{\oplus}:=\bigoplus_{k=0}^{\infty}
(\complex^2)^{\otimes k}.
\end{align}
When we use this space for storing quantum information, 
we cannot determine the length of stored state without state demolition.

Hence the variable-length lossless coding is formulated as follows.
When we compress the quantum state on the $n$-fold tensor product system of ${\cal H}(=\complex^d)$, i.e., ${\cal H}^{\otimes n}$,
the encoder is given by the isometry $U_n$ from ${\cal H}^{\otimes n}$ to ${\cal H}_{\oplus}$,
and the decoder is given as the inverse map from the image $U_n$ to  
${\cal H}^{\otimes n}$.

Note that our definition is more general than Bostr\"{o}m and Felbinger \cite{BF}
because they assume that the compressed state of the basis state belongs
to the space $(\complex^2)^{\otimes n}$, i.e., is not a superposition of 
different-length states.
Further, when we store quantum state by the Fock state, 
we do not have to know the length of the stored state.
Alternatively,
the average energy is an important quantity for physical realization
because a higher energy damages the communication channel and the storage.
Note that the energy is not necessarily to be measured.
In this case, it is natural to treat the Hamiltonian 
$H:=\sum_{k=0}^{\infty} k P_k$, where $P_k$ is the projection to the space 
$(\complex^2)^{\otimes k}$.
In this case, when the initial state is given by $\rho$,
the average energy is given as $\Tr H U \rho U^{\dagger}$.
Hence, it is suitable to consider the minimization of the average energy for 
a given ensemble $\{(\rho_i, p_i)\}$ on ${\cal H}^{\otimes n}$.

For any lossless code $U_n$ on ${\cal H}^{\otimes n}$,
we can choose the density $\sigma(U_n)$ on ${\cal H}^{\otimes n}$ such that 
\begin{align}
- \Tr  \rho \log \sigma(U_n) -
\log \lceil n \log d \rceil
\le \Tr H U \rho U^{\dagger}
\Label{thm2}
\end{align}
for any state $\rho$ on ${\cal H}^{\otimes n}$.
Since 
the relation $- \Tr \rho_p \log \sigma(U_n) -H(\rho_p)= D(\rho_p\|\sigma(U_n))$
holds for the mixture of the ensemble $\rho_p:=
\sum_i p_i \rho_i$,
the average energy $\Tr H U \rho_{p} U^{\dagger}$
is greater than $H(\rho_p)- \log_2 (n \log_2 d)$.

Next, we consider the small class of quantum lossless codes.
When we concatenate two general quantum lossless codes, 
the concatenated lossless code is not necessarily determined.
In order to avoid this problem, we consider the prefix quantum lossless code.
A code $U_n$ on ${\cal H}^{\otimes n}$ is called prefix
when there exists a basis $\{|e_i \rangle\}_{i=1}^{d^n} $ on ${\cal H}^{\otimes n}$ 
(the base $|e_i \rangle$ does not necessarily have the tensor product form.)
such that the state $U_n|e_i \rangle$ has the form $|\phi_1(i) \cdots \phi_k(i)\rangle 
\in (\complex^2)^{\otimes k}$,
and the classical code $\phi(i):=(\phi_1(i) \cdots \phi_k(i))$ satisfies the prefix condition.
Since the classical concatenated code of two classical prefix codes
can be defined,
the concatenated quantum lossless code of two prefix quantum lossless codes
can be defined.
Hence, it is natural to restrict our quantum lossless codes to prefix lossless codes.

As is known in information theory \cite{HK},
any classical prefix code $\phi$ satisfies Kraft inequality
\begin{align*}
\sum_{i}2^{-|\phi(i)|}\le 1,
\end{align*}
where $|\phi(i)|$ is the coding length.
For any prefix quantum lossless code $U_n$,
we choose a classical prefix code $\phi$ such that 
$U_n|e_i\rangle=|\phi(i)\rangle$.
Then, a quantum version of Kraft inequality 
\begin{align}
\Tr 
\sum_i 
2^{-\langle i|U_n^{\dagger} H U_n |i\rangle }
|e_i\rangle \langle e_i|
=\sum_{i}2^{-|\phi(i)|}
\le 1 \Label{6-4-1}
\end{align}
holds 

Conversely, 
as is known in information theory \cite{HK},
for any probability distribution $p_i$,
there exists a prefix classical code $\phi$ such that
\begin{align}
-\log p_i \le |\phi(i)| \le -\log p_i +1.\Label{6-3-4}
\end{align}
Hence, for a given state $\sigma$ on ${\cal H}^{\otimes n}$,
we focus on its diagonalization 
$\sigma=\sum_i p_i |i\rangle \langle i|$.
Then, we choose the prefix classical code $\phi$ satisfying 
(\ref{6-3-4}).
Then, we can define the prefix quantum lossless code $U_n$ by
\begin{align*}
U_n |e_i\rangle = |\phi(i)\rangle  .
\end{align*}
The prefix quantum lossless code $U_n$ satisfies that
\begin{align*}
-\Tr \rho \log \sigma 
\le 
\Tr H U \rho U^{\dagger}
\le -\Tr \rho \log \sigma +1
\end{align*}
for any state $\rho$ on ${\cal H}^{\otimes n}$.
That is, a prefix quantum lossless code almost corresponds to a quantum state.
Therefore, we can identify a prefix quantum lossless code with the corresponding quantum state.
Since the relation $- \Tr \rho_p \log \sigma(U_n) -H(\sigma_p)= D(\rho_p\|\sigma(U_n))$
holds for the mixture of the ensemble $\rho_p$,
the minimum of
the average energy $\Tr H U \rho_{p} U^{\dagger}$
among prefix quantum lossless codes
is almost equal to the von Neumann entropy $H(\rho_p)$.
Further, the inequality (\ref{thm2}) guarantees that 
if we remove the prefix condition, 
it is impossible to improve the average energy so much.

When we use a prefix quantum lossless code corresponding to a state $\sigma$
and the true mixture state is $\rho_p$,
the average energy is
\begin{align*}
-\Tr \rho_p \log \sigma = H(\rho_p)+ D(\rho_p\|\sigma).
\end{align*}
Therefore, the relative entropy $D(\rho_p\|\sigma)$ can be regarded as the redundancy of
the quantum lossless code $\sigma$ with respect to $\rho_p$.

When the true mixture is $\rho^{\otimes n}$
and the prefix quantum lossless code $\sigma_{P_n,n}$ defined in Section \ref{s3} is applied,
the asymptotic redundancy is given by (\ref{6-3-1}).
Hence, the mini-max 
asymptotic redundancy is $\frac{d^2-1}{2} \log n + C_d + \log \int_{Y^d} e^{C(\vec{p})} d \vec{p}$,
which is attained when 
we choose the mini-max code $\sigma_{J,n}$ or 
$\tilde{\sigma}_{J,n}$.
Therefore, the prefix quantum lossless code $\sigma_{J,n}$ and $\tilde{\sigma}_{J,n}$
can be used as universal quantum lossless data compression.

\section{Discussion}
We have found a sequence of states $\{\sigma_n\}$ such that 
the relative entropy $D(\rho^{\otimes n}\| \sigma_n)$ behaves as 
$\frac{d^2-1}{2}\log n$ universally. While this result known only in qubit case,
the general case had been open for more than ten years.
In this derivation, Matsumoto and Hayashi \cite{MaHa07}'s calculation plays an essential role.
Further, we have solved the asymptotic mini-max problem concerning $D(\rho^{\otimes n}\| \sigma_n)- \frac{d^2-1}{2}\log n$.
It has been checked that our optimal value better than Krattenthaler and Slater \cite{KS}'s result in the qubit case.
Our discussion is different from the original Clarke and Barron\cite{CB,CB2}'s discussion because they consider the optimization among mixtures of $n$ i.i.d. distributions.
Our method can be translated to the classical case when the family is given as the full multinomial distribution family.
In such a case, the derivation of mini-max problem 
can be expected to be shorter than the original derivation\cite{CB,CB2}.

In addition, we have revisited quantum variable-length lossless data compression, 
and characterized the average energy (length) by (\ref{thm2}) for the general case.
This result (\ref{thm2}) has a similar form as Theorem 5 in \cite{CC}, 
but is a little different from their result.
In addition, our definition of quantum lossless code is more general.
Then, our result is stronger. 
While they used the method of majorization, we do not use it.

Concerning the prefix case,
we have derived the correspondence between the prefix quantum lossless code 
and a density matrix through a quantum version of Kraft inequality (\ref{6-4-1}).
Using this relation, we have applied our result (\ref{P2}) and (\ref{6-2-3}) to 
quantum variable-length lossless data compression.
It has been clarified that 
it is possible to compress multi-copy state universally by lossless code 
$\sigma_{J,n}$ or
$\tilde{\sigma}_{J,n}$.

Further, 
the relation between obtained approximation and 
other universal protocols are not discussed in this paper.
This relation will be treated in a future paper\cite{prep}.
Obtained universal approximation may be applied to other topic in quantum information \cite{H-book}.
Hence, it is interesting to discuss this application.

\section*{Acknowledgment}
The author thanks Professor Hiroshi Nagaoka for explaining Clarke and Barron's result\cite{CB,CB2}. He is also grateful to Professor Keiji Matsumoto for helpful discussion concerning representation theory.
This research was partially supported by a Grant-in-Aid for Scientific Research on Priority Area `Deepening and Expansion of Statistical Mechanical Informatics (DEX-SMI)', No. 18079014 and
a MEXT Grant-in-Aid for Young Scientists (A) No. 20686026.

\appendix
\section{Proof of (\ref{thm2})}
Let $P_k$ and $A_0$ be 
the projections to the space $\complex^2$
and the image of the code $U_n$, respectively.
We denote the projection to the range of operator $X$ by $P(X)$.
Now, we define the projection 
$A_k:= A_{k-1}-P(A_{k-1} P_k A_{k-1})$ inductively.
Since the projection $P(A_{k-1} P_k A_{k-1})$ is commutative with $A_{k-1}$,
$A_k$ is a projection.
Since the rank of $A$ is finite,
there exists a number $K$ such that $A_k=0$ for $k \ge K$.
Thus, $A_0=\oplus_{k=1}^{K} P(A_{k-1} P_k A_{k-1})$.
Define the unitary $V$ and the projection $B$
\begin{align*} 
P(A_{k-1} P_k A_{k-1})
=&V P( P_k A_{k-1} P_k) V^\dagger \\
B:=& \oplus_{k=1}^{K} P( P_k A_{k-1} P_k).
\end{align*}
Then, $A_0=V B V^\dagger $.
Since 
all non-zero eigenvalues of $P(A_{k-1} P_k A_{k-1}) H P(A_{k-1} P_k A_{k-1})$
are greater than $k$,
$
P(A_{k-1} P_k A_{k-1}) H P(A_{k-1} P_k A_{k-1})
\ge 
V P( P_k A_{k-1} P_k) 
H P( P_k A_{k-1} P_k) V^\dagger $.
Thus, $A_0 H A_0 \ge V B H BV^\dagger$.

Now, we construct a state $\sigma (U_n)$.
We choose a basis $\{|f_i\rangle\}$ on the range of $B$ such that
$|f_i\rangle \langle f_i|$ is commutative with all of $P( P_k A_{k-1} P_k)$.
Using the basis $|e_i\rangle:= U_n^\dagger V |f_i\rangle $,
we define the matrix $\tilde{\sigma} (U_n) := 
\sum_i 2^{-\langle f_i|H | f_i\rangle}|e_i\rangle\langle e_i|$.
The average length satisfies that
\begin{align*}
& \Tr  U_n \rho U_n^{\dagger} H
=  
\Tr  U_n \rho U_n^{\dagger} 
A_0 H A_0 \\
\ge &
\Tr  U_n \rho U_n^{\dagger} 
V B H BV^\dagger
=
-\Tr \rho \log \tilde{\sigma} (U_n) 
\end{align*}
for any state $\rho$.
This inequality guarantees that it is sufficient to treat the code $V^\dagger U_n$ instead of $U_n$.

Next, we calculate $\Tr \tilde{\sigma} (U_n)$.
we reorder $|e_i\rangle$ with the order 
increasing $\langle f_i|H|f_i\rangle$.
When $M=2^1+2^2+2^3+\cdots +2^m (\ge 2^m)$,
we have
\begin{align*}
\sum_{i=1}^{M} 2^{-\langle f_i|H|f_i\rangle} \le m \le \log M.
\end{align*}
Thus, 
\begin{align*}
\sum_{i=1}^{M'} 2^{-\langle f_i|H|f_i\rangle} \le \lceil\log M'\rceil
\end{align*}
for any $M$.
Since $\dim {\cal H}^{\otimes n}=d^n$, 
\begin{align*}
\Tr \tilde{\sigma} (U_n) \le \lceil n \log d \rceil. 
\end{align*}
Defining the state $\sigma (U_n):= \frac{1}{\Tr \tilde{\sigma} (U_n)}\tilde{\sigma} (U_n)$,
we obtain 
\begin{align*}
\Tr  U_n \rho U_n^{\dagger} H
\ge 
-\Tr \rho \log \sigma (U_n) 
- \log \lceil n \log d \rceil
\end{align*}
for any state $\rho$.

\end{document}